\documentclass[conferenece]{IEEEtran}
\usepackage{graphicx}
\usepackage{subfig}
\usepackage{cuted}
\usepackage{amssymb, booktabs, array}
\usepackage{multirow,float}
\usepackage{bbding}
\usepackage{makecell}
\usepackage{booktabs}
\usepackage{times,soul,color}
\usepackage{color}
\usepackage{adjustbox}
 \usepackage{bbm}
\usepackage{caption}
\usepackage[usestackEOL]{stackengine}
\usepackage[compress]{cite}
\usepackage[switch,pagewise]{lineno} 
\usepackage{setspace}%
\usepackage{url}
\usepackage{enumitem}
\usepackage[utf8]{inputenc}
\usepackage[english]{babel}

\usepackage{breakurl}
\usepackage[cmintegrals]{newtxmath} 

\definecolor{Hilight}{rgb}{0,0,0} 
\definecolor{Rlight}{rgb}{1,0,0}{\large }

\sethlcolor{Hilight}
\sethlcolor{Hilight} \sethlcolor{Blight}

\usepackage{algorithm}
\usepackage[noend]{algpseudocode}
\usepackage{multirow}
\usepackage{fancyhdr}

\usepackage{xcolor,colortbl}
\definecolor{Gray}{gray}{0.85}

\newcolumntype{g}{>{\columncolor{Gray}}c}

\begin{document}

\title{Robust Proximity Detection using On-Device Gait Monitoring}

\author{Yuqian Hu, Guozhen Zhu, Beibei Wang, and K. J. Ray Liu\\
Origin Research, Rockville, MD, 20852\\
E-mail: \{yuqian.hu, guozhen.zhu, beibei.wang,  ray.liu\}@originwirelessai.com}





\maketitle

\begin{abstract}
Proximity detection in indoor environments based on WiFi signals has gained significant attention in recent years. Existing works rely on the dynamic signal reflections and their extracted features are dependent on motion strength. To address this issue, we design a robust WiFi-based proximity detector by considering gait monitoring. Specifically, we propose a gait score that accurately evaluates gait presence by leveraging the speed estimated from the autocorrelation function (ACF) of channel state information (CSI). By combining this gait score with a proximity feature, our approach effectively distinguishes different transition patterns, enabling more reliable proximity detection. In addition, to enhance the stability of the detection process, we employ a state machine and extract temporal information, ensuring continuous proximity detection even during subtle movements. Extensive experiments conducted in different environments demonstrate an overall detection rate of 92.5\% and a low false alarm rate of 1.12\% with a delay of 0.825s.

\end{abstract}


\section{Introduction}

The demand for human proximity detection has grown significantly across diverse indoor applications. Human proximity detection refers to the ability to identify and monitor the presence and proximity of individuals in relation to specific areas, objects, or systems. This technology plays a crucial role in home automation, indoor localization, security, and efficiency enhancement in various contexts. While popular vision-based technologies like cameras \cite{singla2014motion, sun2022indoor}, infrared\cite{jacob2022intrusion, yang2015pir}, lidar \cite{Wang_2022_CVPR}, and radar \cite{Gurbuz2019radar,li2017radar} are commonly used for human motion detection and localization, they can be privacy-invasive or require specialized sensors for accurate measurements. In cases where additional sensors are unavailable or privacy preservation is a concern, alternative solutions are needed.


In today's IoT-driven world, the popularity of IoT devices has surged, and the widespread use of WiFi connectivity in these devices has led to the realization that wireless sensing will play a major role in IoT applications. By utilizing WiFi radio interactions during signal propagation, wireless devices can passively detect human movements and activities through the analysis of wireless signals \cite{chenshu2022wifidomore, ray2019wirelessai}. This approach makes passive indoor proximity detection more realistic and feasible. 
Instead of relying on invasive vision-based technologies, analyzing the wireless signal propagation allows for non-intrusive detection of human presence and movements. This advancement in WiFi-based proximity detection enables a more privacy-friendly and accessible solution for practical use.

However, implementing a robust proximity detector through on-device WiFi signals is not a straightforward task. The ability of WiFi signals to penetrate walls and other obstacles helps with whole-home coverage, but it hinders proximity detection.
Many existing research works have studied the utilization of WiFi signals for motion detection, particularly by leveraging fine-grained physical layer channel state information (CSI) \cite{gong2015wifi, lv2018sied, gu2017motion, zhang2019widetect}, but they primarily focus on motion detection within a wide coverage area, overlooking the specific aspect of proximity detection. 
Other works concentrate on localization, but they often require complex setups involving multiple transceivers with specific geometric arrangements or dedicated calibration and training for precise localization \cite{xiao2013pilot,zhou2017device, qian2018widar2,shi2018fingerprinting,zhang2018defi}. Such complexities are unnecessary for proximity detection.

In our prior works \cite{yuqian2022proximityIOTJ, yuqian2021icasspproximity}, we delved into the underlying multipath propagation properties and proposed a proximity feature based on the correlation of adjacent subcarriers to differentiate between nearby and faraway motion. This feature has proven to be sensitive to the distance between a moving human and the devices. However, due to its reliance on dynamic reflections from the human target, the proximity feature is highly affected by the strength of motion.
For instance, when a person changes from larger movements such as walking to smaller movements such as reading after approaching the devices, the value of the proximity feature drops, causing the proximity detector to fail in continuously detecting the person in close proximity.
Therefore, a more stable proximity detection is still in need.

Recognizing that humans typically approach or move away from devices by walking, we consider gait information as supplementary factor to support human proximity detection.
Gait can be inferred from the speed, which is not easily estimated from WiFi signals. Inspired by previous WiSpeed work \cite{zhang2018wispeed}, we estimate the speed from the autocorrelation function (ACF) of CSI based on a statistical model. With the ACF spectrum and speed information available, we propose a gait score, which serves as an indicator of the presence of gait.

By further combining the proximity feature and gait information, we design a state machine that accumulates temporary information and triggers or untriggers proximity detection only when transition states, such as approaching or leaving, are detected. In this way, even if the detected human performs subtle movements after being initially detected, the proximity detection continues until the leaving state is identified.


\begin{figure*}[t]
  \centering
  \includegraphics[width=0.87\linewidth]{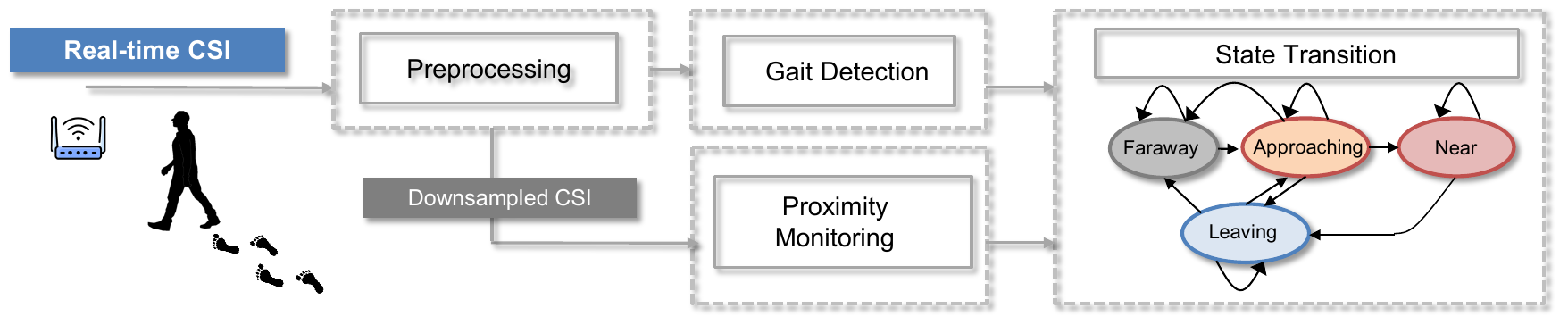}
    \caption{System overflow.}
    \label{fig:system_overflow}
\end{figure*}

In summary, the main contributions of this paper are as follow:
\begin{itemize}
    \item We develop a novel gait score that leverages the ACF of CSI and the estimated speed to accurately evaluate the presence of gait. By combining this gait score with the proximity feature, we can successfully distinguish different transition patterns for proximity detection.
    \item We extract temporal information and implement a state machine to effectively distinguish between nearby and faraway human. This approach ensures continuous proximity detection even during subtle movements, enhancing the stability of the detection process.
    
    \item Through extensive experiments in diverse environments, we validate the effectiveness of our proposed system. It achieves a high overall detection rate while maintaining a low false alarm rate across various scenarios.
\end{itemize}

The rest of the paper is organized as follows. Section \ref{sec:system_design} presents the design of the proposed system and the fundamental of each module is explained. Section \ref{sec:evaluation} offers a thorough examination of the experiments and evaluations conducted. We discuss the future work in Section \ref{sec:future_work}. Finally, the paper concludes in Section \ref{sec:conclusion}.

\section{System Design}
\label{sec:system_design} 

\subsection{CSI Primer} 
In wireless communication, CSI (a.k.a channel frequency response or CFR), describes the propagation of the WiFi signals from a transmitter (Tx) to a receiver (Rx). The CSI estimated over a subcarrier with frequency $f$ at time $t$ can be represented as
\begin{equation} \label{eqn:csi_definition}
H(t,f) = \frac{Y(t,f)}{X(t,f)}, \\
\end{equation}
where $X(t,f)$ and $Y (t, f )$ denote the transmitted signal by the Tx and the received signal at the Rx, respectively. Since the transmitted WiFi signals experience multiple reflections in their propagation in indoor environments, the corresponding CSI can be written as a collection of radio propagation along different paths:
\begin{equation}
    H(t,f) = \sum^{L}_{l = 1} \alpha_l(t) e^{-j2\pi f\tau_l(t)},
    \label{eqn: multipath_channel_CSI}
\end{equation}
where $\alpha_l$ is the multipath coefficients of the $l$-th independent multipath component and $\tau_l$ is the associated propagation delay. 
 In practice, since the imperfect synchronization of the commercial WiFi devices often results in random noise in the CSI phase that is difficult to be efficiently cleaned \cite{chen2016achieving,xie2015precisePDP}, researchers usually rely  on the more reliable CSI amplitude \cite{zhang2019widetect, li2020wiborder}, which can be measured through the power response $G(t,f)$ as 
\begin{equation}
G(t,f) \triangleq |H(t,f)|^2.
\label{eqn:G_definition}
\end{equation}

\subsection{System Design}

As depicted in Fig.  \ref{fig:system_overflow}, the proposed system is composed of \emph{CSI Preprocessing}, \emph{Gait Detection}, \emph{Proximity Monitoring} and \emph{State Transition}. 

\subsubsection{CSI Preprocessing}

To mitigate the impact of the automatic gain controller on the CSI amplitude, the measured CSI power response $G(t,f)$ are first normalized followed by a Hampel filter \cite{Davies1993hampel}, to remove the outliers in time domain.

\subsubsection{Proximity Monitoring}
\begin{figure}[t]
  \centering
  \includegraphics[width=0.9\linewidth]{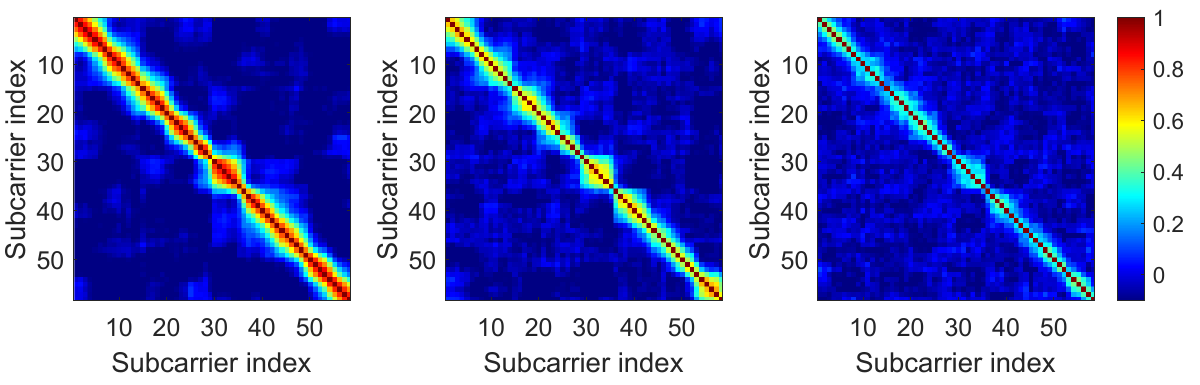}
    \caption{Correlation matrix among subcarriers at human-device distances of 1m, 3m, and 5m.}
    \label{fig:proximity_feature}
\end{figure}
Our previous study has shown that the correlation of CSI power response on adjacent subcarriers is an effective indicator of the rough distance between the moving human and the WiFi devices \cite{yuqian2022proximityIOTJ, yuqian2021icasspproximity}. As illustrated in Fig. \ref{fig:proximity_feature}, the closer the human motion to the WiFi device, the higher correlation between the CSI power response on adjacent subcarriers.  
Therefore, a proximity feature can be defined as 
\begin{equation}
    F_{p} = \frac{1}{N_s-1}\sum^{N_s-1}_{n = 1}\rho_G (f_n,f_{n+1}),
    \label{eqn:proximity_score_def}
\end{equation}
 where $\rho_G (f_n, f_{n+1})$ is the correlation over adjacent subcarriers $n$ and $n+1$, and $f_n$ and $f_{n+1}$ are the corresponding subcarrier frequencies.

In addition, we also utilize the slope ($F_s$) of the proximity feature $F_p$ within a certain time window in the proposed proximity monitoring. As a human subject moves closer to or farther away from the device, the value of $F_p$ correspondingly increases or decreases. Consequently, the slope $F_s$ captures the informative trend of $F_p$, enabling us to gain valuable insights into the dynamics of proximity.

\subsubsection{Gait Detection}
\begin{figure}[t]
  \centering
\includegraphics[width=0.9\linewidth]{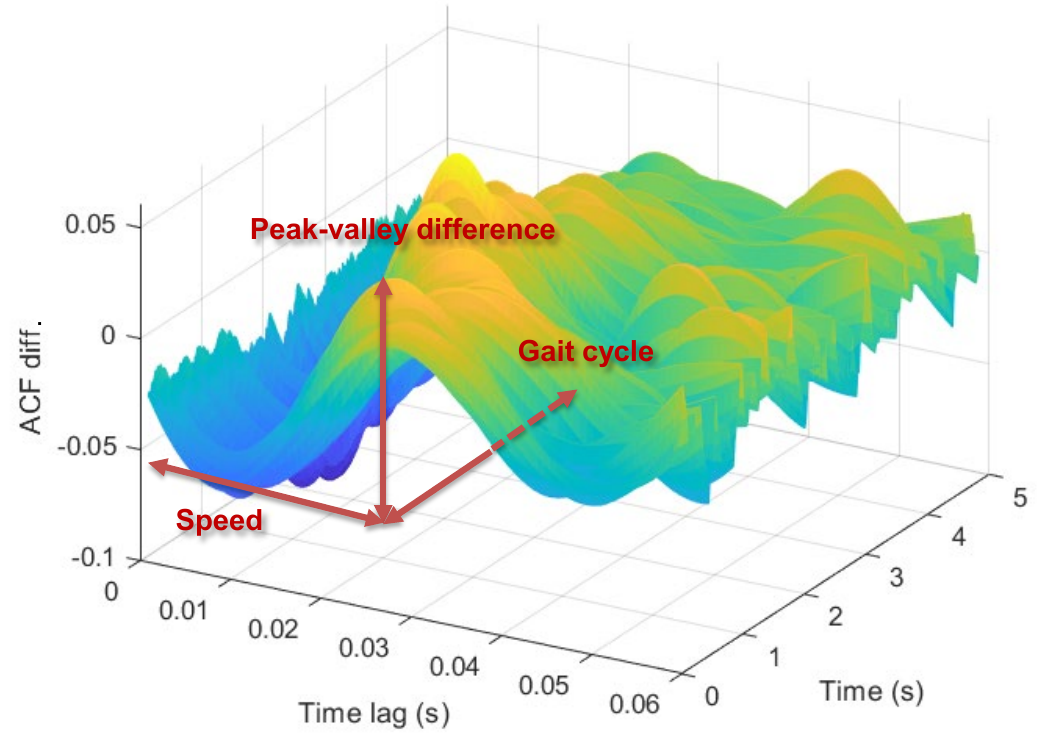}
    \caption{Gait information from the differential of ACF.}
    \label{fig:gait_3dacf}
\end{figure}
As the proximity feature relies on the dynamic reflection of WiFi signals, it is susceptible to being impacted by the intensity of motion, and smaller movements in the same target area may result in smaller feature values compared to larger movements \cite{yuqian2022proximityIOTJ}. If the user enters the target area and remains still, the proximity features will not indicate the nearby motion in a consistent way. Since the change of proximity involves walking motion, gait information can be used to determine if a human subject has entered/left a specific area. 

The existence of human gait can be detected by identifying gait cycles. A gait cycle is a walking pattern, starting from the moment one foot striking the ground to the moment that same foot strikes the ground again. It can be identified by the speed during walking, because each step has a speed peak during mid-stance, followed by a decrease in speed as the foot leaves the ground, and an increase in speed during the swing phase until the foot strikes the ground again
 \cite{chenshu2022gaitway, zahid2022gaitcube}. 
 The walking speed can be reliably estimated from a statistical model in our previous study \cite{zhang2018wispeed, yuqian2022defall}, based on the ACF of CSI power as
\begin{equation}
    \begin{aligned}
    R_G(f, \triangle t) = g(f)J_0(kv\triangle t),
    \end{aligned}
    \label{eqn:acf_speed}
\end{equation}
where $J_0$ is the zero-order Bessel function, $k$ is the wave number, $f$ is the subcarrier frequency and $g(f)$ is gain coefficient on that subcarrier. 
Then, we can further define a gait score which reflects the probability of gait existence, based on the following consideration as illustrated by Fig. \ref{fig:gait_3dacf}.

\begin{itemize}
    \item In the time domain, the fluctuation of the estimated speed can be used to infer the human gait cycle. 
    \item The confidence of the gait speed can be extracted based on the distribution of normal human walking speed.
    \item The significance of the peak and valley also indicates the confidence of the speed estimation. Therefore, we use the peak-valley difference as the weights to calculate the gait score. 
\end{itemize} 
 
Overall, the gait score is formulated as
\begin{equation}
\begin{aligned}
    F_g &=  w \cdot p(v) \mathbbm{1}_{(0, \infty)} (w) \mathbbm{1}_{[0.5, 1.5]} (c),
\end{aligned}
\end{equation}
where $w$ is the difference between the prominence of peak and valley of ACF differential. $w$ equals zero if there is no peak or valley detected. $c$ is the number of gait cycle per second, and is confined to be between 0.5 and 1.5 for a normal walking. $p(v)$ is the probability that the observed speed would have occurred if the speed is from human walking. Since the pedestrian walking speed follows a normal distribution with an estimated mean of 1.34 m/s and a standard deviation of 0.37 m/s \cite{Weidmann2006walkingspeed}, we can calculate the probability as:
\begin{equation}
    p(v) = 1 - 2|\Phi(\frac{v-1.34}{0.37}) - 0.5|,
    \label{eqn:gait_score_def}
\end{equation} 
where $\Phi(\frac{v-1.34}{0.37})$ is the cumulative distribution function (CDF) of a non-standard normal distribution with mean $1.34$ and standard deviation $0.37$.

\subsubsection{State Transition}
\begin{figure}[t]
  \centering
  \includegraphics[width=0.95\linewidth]{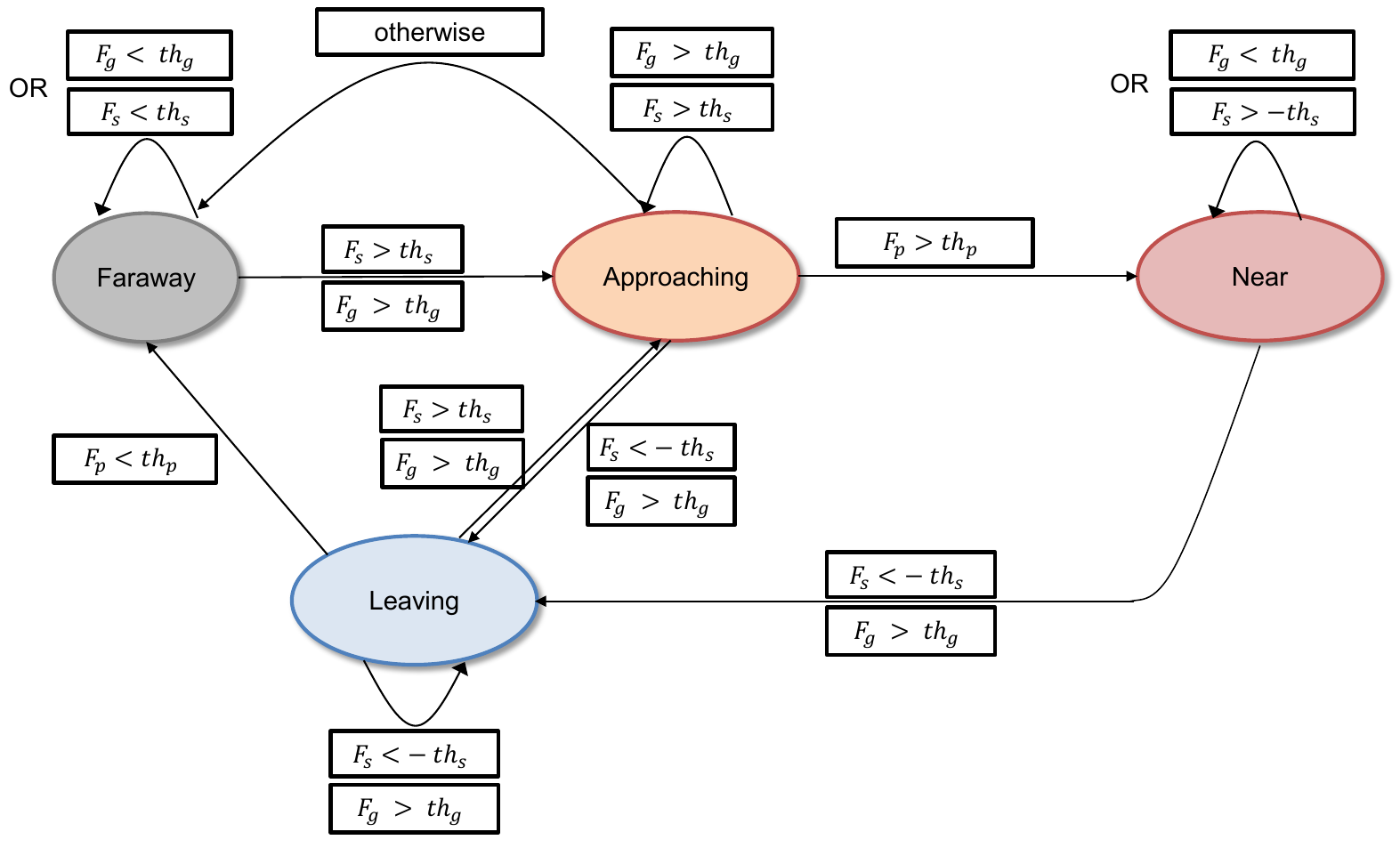}
    \caption{Illustration of FSM for proximity monitoring.}
    \label{fig:statemachine_flowchart}
\end{figure}

The combination of both proximity features and gait information allows one to infer whether a moving human has entered the proximate area. To this end, we propose a finite-state-machine (FSM) consisting of multiple states as Fig. \ref{fig:statemachine_flowchart} shows. By monitoring the trend of the proximity feature and gait existence, the FSM enhances the accuracy of the inference process. It contains four states: Faraway, Approaching, Near, and Leaving, with details as follows.

\begin{figure*}[t]
  \centering
  \includegraphics[width=0.9\linewidth]{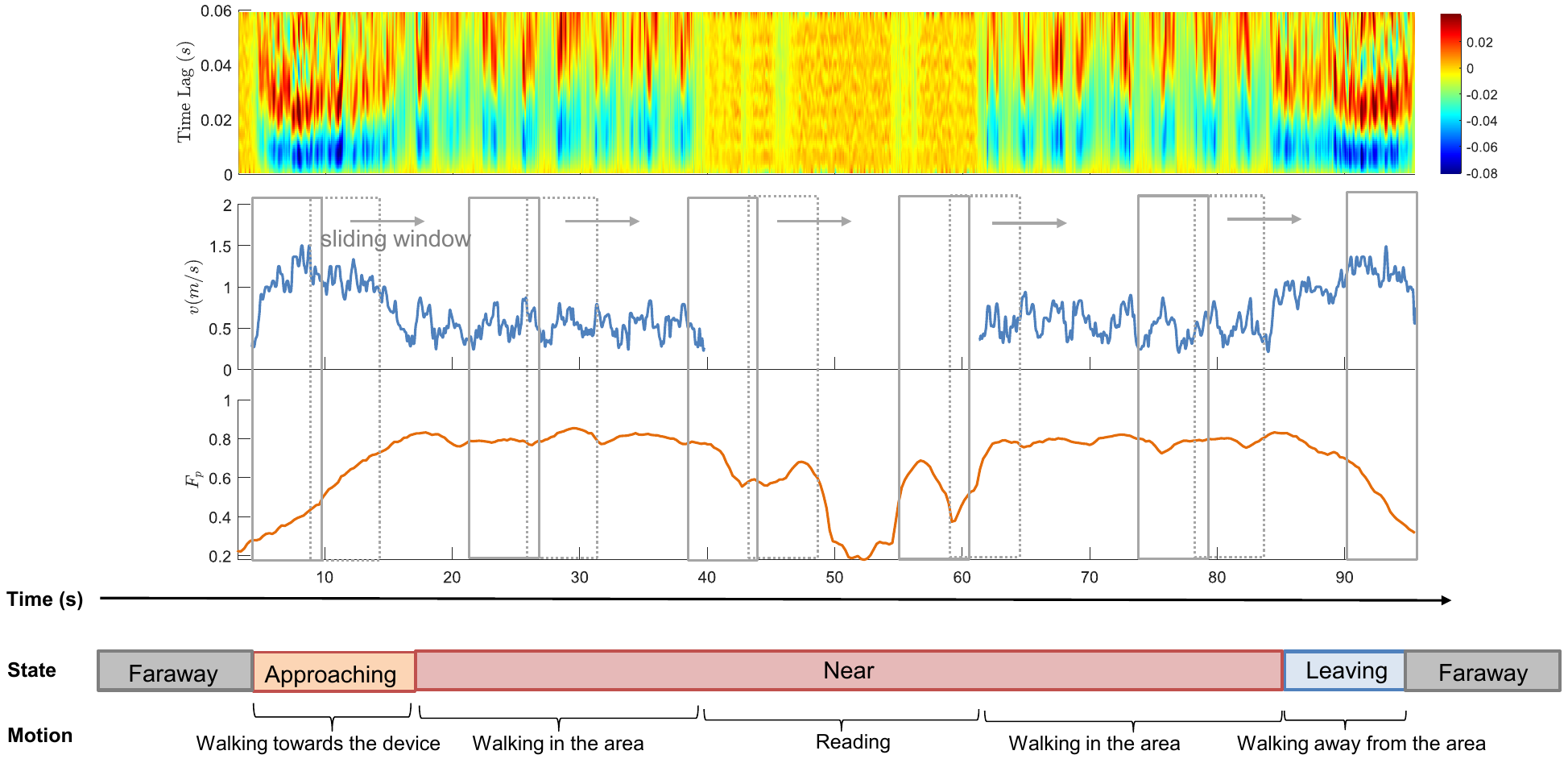}
    \caption{The illustration of a whole process with four states. }
    \label{fig:state_example}
\end{figure*}

\begin{itemize}
    \item The Faraway state is the system's default state, in which the detector continuously monitors the ACF spectrum and calculates the proximity feature while outputting the possibility of detecting human gait. If the system detects an increase in proximity feature along with human gait, it transitions to the Approaching state. Otherwise, it remains in the Faraway state.
    \item The FSM enters the Approaching state with the presence of human gait and an increasing in the proximity feature. The system keeps checking the proximity feature and gait. If the proximity feature reaches a threshold indicating proximity, the FSM switches to the Near state. Otherwise, it remains in the Approaching state. In the absence of gait detection and the high proximity feature, the FSM switches back to the Faraway state.
    \item In the Near state, the human subject has entered the pre-defined proximate area, irrespective of the intensity of motion. Based on an assumption that the human will leave only by walking, if the system detects a decrease in proximity feature and human gait simultaneously, then the FSM transitions to the Leaving state. Otherwise, the system remains in the Near state. 
    \item The Leaving state denotes that the human is walking out of the proximate area. When the proximity feature decreases to a threshold indicating departure, the FSM switches back to the default initialization state.
\end{itemize}

Note that accurate gait extraction from CSI requires a high sounding rate \cite{chenshu2022gaitway} while the proximity feature does not \cite{yuqian2022proximityIOTJ}. Therefore, we have incorporated a downsampling module into the proposed system, as depicted in Fig. \ref{fig:system_overflow}. This module serves to reduce the computational overhead, making the system more practical for real-time applications.

Fig. \ref{fig:state_example} provides an illustrative example of how the extracted features evolve over time as a person approaches a device. Initially, as the person walks towards the device, noticeable fluctuations appear in the spectrum of ACF differential, as well as in the estimated speed ($v$), accompanied by an increasing value in the proximity feature $F_p$. Once the person enters the target area, $F_p$ remains high, but it drops if the person engages in small motions (e.g. reading). Despite the drop in $F_p$, the state machine remains in the Near state because there is no observed gait pattern. Only upon a gait pattern detection and a decrease in $F_p$, the Leaving state is triggered. The integration of gait features with proximity features within the state machine effectively enhances system robustness when the human is in a quasi-static status.

\section{Evaluation}
\label{sec:evaluation}
\subsection{Hardware Platform}
We build one pair of Tx-Rx prototypes equipped with off-the-shelf WiFi cards. Each prototype has 2 omni-directional antennas. The center frequency is configured as 5.18 GHz with a bandwidth of 40 MHz. The sounding rate is set to be 1500Hz and downsampled to 30Hz for the proximity monitoring module for lowering computation complexity.

\begin{figure*}[t]
\subfloat[Environment 1.]{
  \centering
  \includegraphics[width=0.31\linewidth]{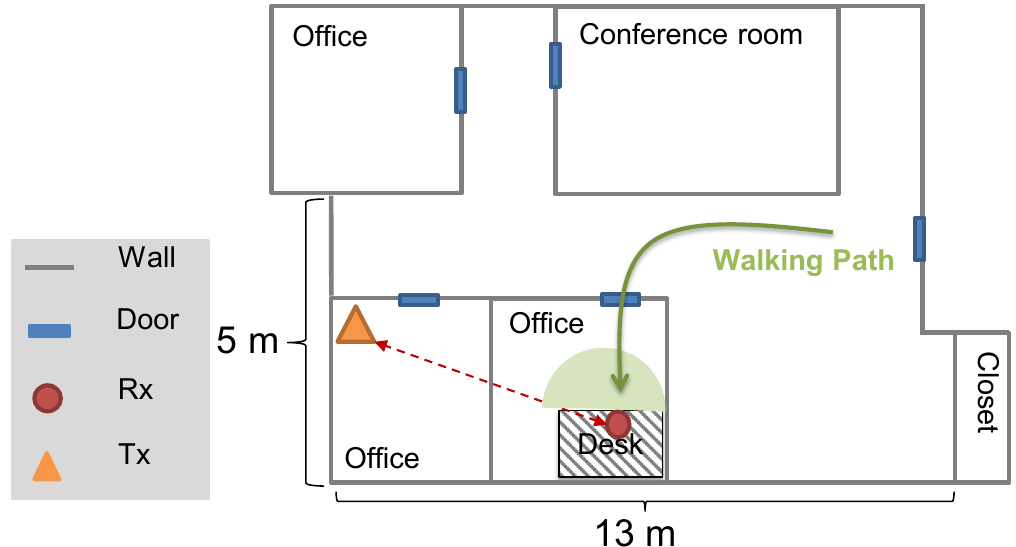}
  \label{fig:floorplan1}
}
\subfloat[Environment 2.]{
  \centering
  \includegraphics[width=0.198\linewidth]{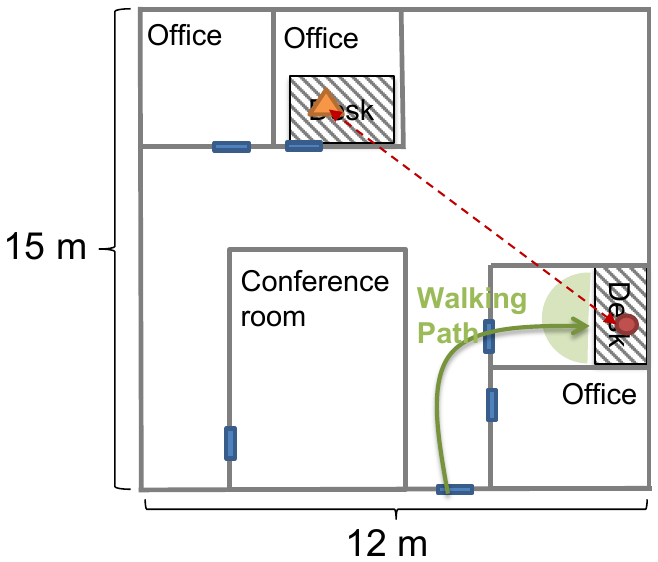}
  \label{fig:floorplan2}
}
\subfloat[Environment 3.]{
  \centering
  \includegraphics[width=0.195\linewidth]{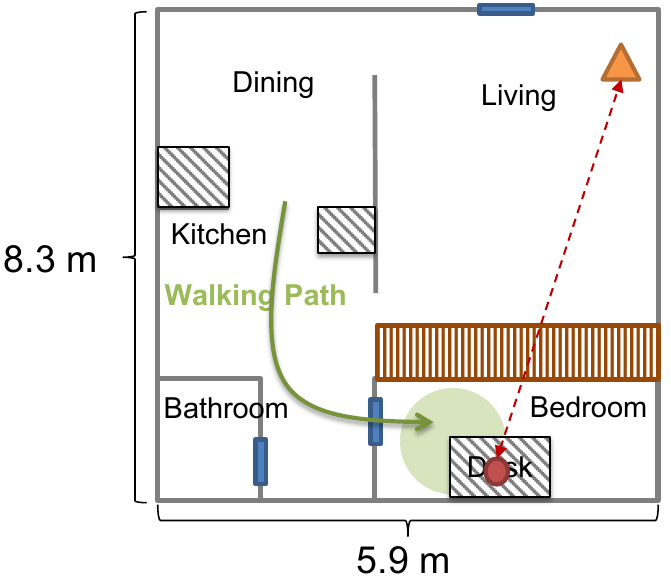}
  \label{fig:floorplan3}
}
\subfloat[Environment 4.]{
  \centering
  \includegraphics[width=0.24\linewidth]{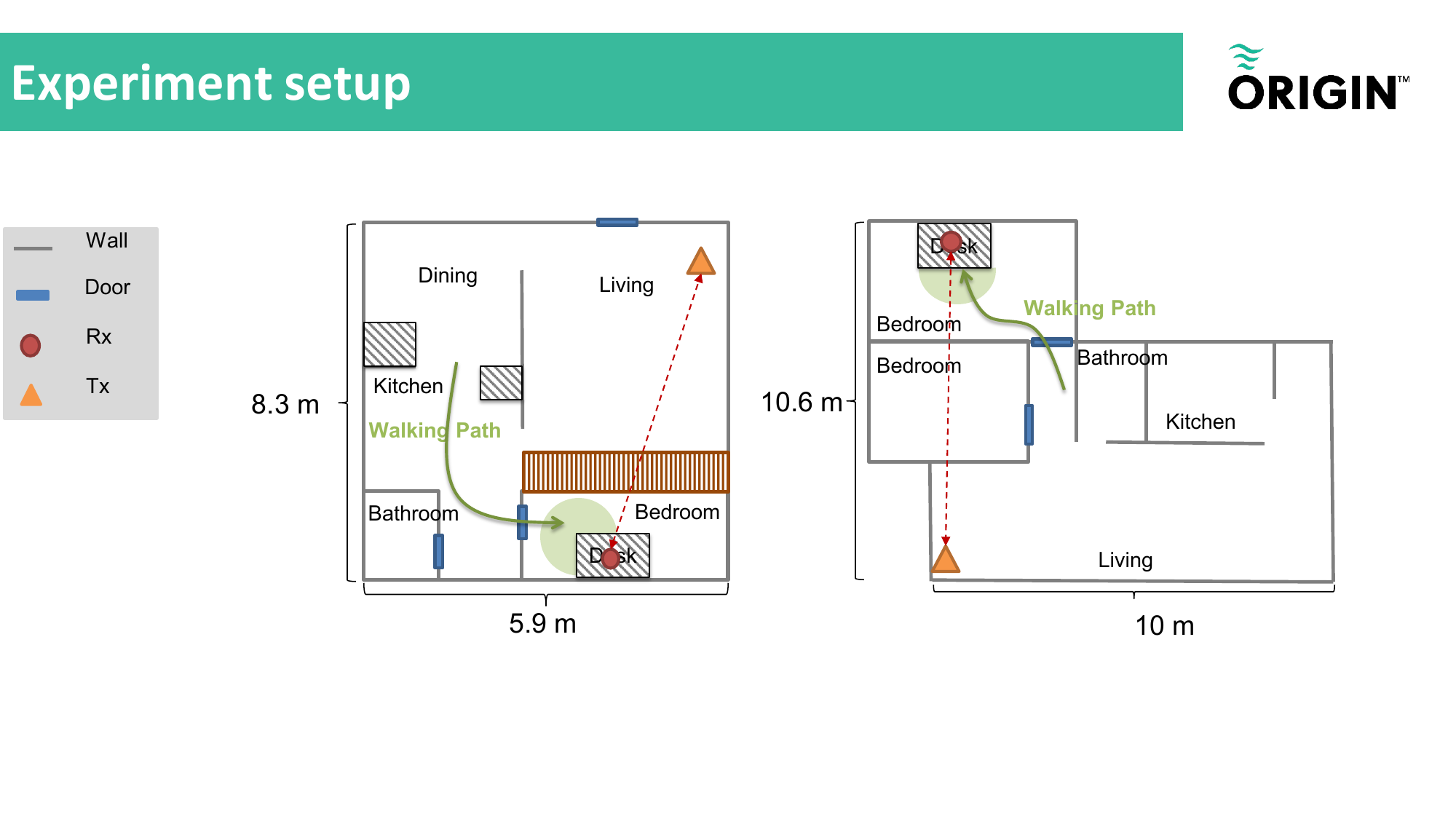}
  \label{fig:floorplan3}
}
    \caption{The experiments environments and deployments.}
    \label{fig:floorplans}
\end{figure*} 

\subsection{Experiments}

To evaluate the performance of our proposed system, we conducted extensive experiments in diverse environments, including two typical office environments and two home environments (one apartment and one single-family house), as depicted in Fig. \ref{fig:floorplans}.
During the experiments, participants walk toward a target device at a normal walking speed and stay in the immediate vicinity of the target device with minor movements (e.g. sitting and reading). Afterwards, they return to the starting point. The sequence of the above-mentioned motion, which can last for a minute or longer, constitute an ``event sample". To assess the false alarm rate, we maintained the devices at the same fixed locations while allowing participants to freely move away from the devices. Note that for each test, there is only a single person in the environment. The ground truth is recorded by timestamp cameras. 

\subsection{Evaluation metrics}
\begin{figure}[tb]
\centering
\subfloat[Aligned duration-based evaluation metric.]{
\includegraphics[width=0.9\linewidth]{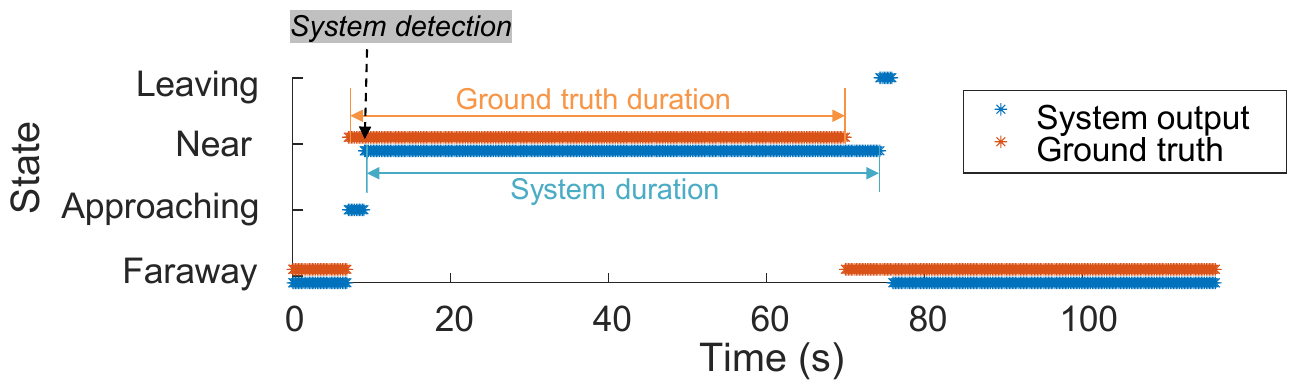}
    \label{fig:instance_duration_metric}
    }\\
\subfloat[Responsiveness evaluation metric.]{
  \includegraphics[width=0.49\linewidth]{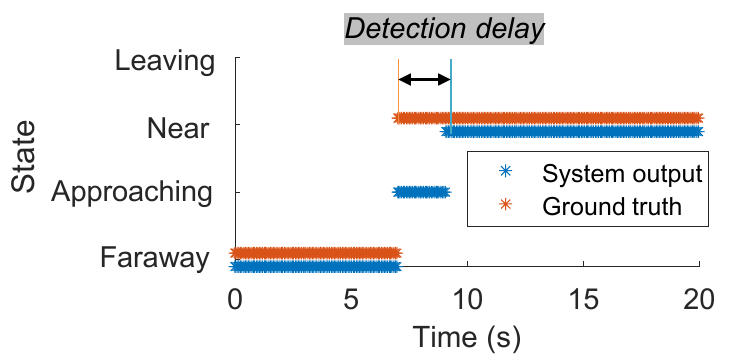}
  \label{fig:delay_metric}
}
\subfloat[False alarm evaluation metric.]{
  \includegraphics[width=0.49\linewidth]{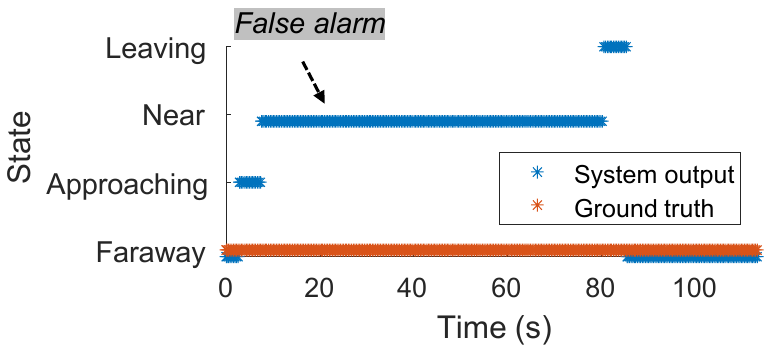}
  \label{fig:false_alarm_metric}
}
    \caption{Definitions of the evaluation metrics. }
    \label{fig:evaluation_metric}
\end{figure}

We use the following metrics to evaluate the performance of the proposed system.
\begin{itemize}
    \item Instance-based accuracy (IA). The IA measures the accuracy of detecting the aforementioned ``event samples'' involving the sequence of an object approaching, entering the proximate area, and leaving it. It is calculated as the ratio of system detections to the ground truth occurrences of this sequence. In mathematical terms, we can write IA as
    \begin{equation}
        IA = \frac{N_{sys}}{N_{GT}},
    \end{equation}
   where $N_{GT}$ represents the actual number of occurrences of the event according to the ground truth, while $N_{sys}$ represents the number of detections made by the system for the same event as Fig. \ref{fig:instance_duration_metric} indicates.  
    \item Duration-based accuracy (DA). We define the proximate area as the area within 1.5 meters of the target device. Using a timestamp camera and the labels on the ground, we can determine the actual duration from a person entering the target area until he/she leaves the area, thereby obtaining a duration-based metric. 
    The duration-based accuracy can be mathematically formulated as:
    \begin{equation}
        DA = \lceil \frac{T_{sys}}{T_{GT}} \rceil^1,
    \end{equation}
    where $T_{sys}$ is the duration of system output staying at the Near state while $T_{GT}$ is the duration of the true Near state. Fig. \ref{fig:instance_duration_metric} illustrates the calculation of DA. Note that the DA can only be calculated on the samples that have both Approaching and Leaving successfully detected.
    
    \item Responsiveness. As illustrated by Fig. \ref{fig:delay_metric}, the responsiveness of the system can be evaluated by measuring the time difference between the system detection and the ground truth, i.e.,
    \begin{equation}
        \tau = t_{GT} - t_{sys},
    \end{equation}
    where $t_{GT}$ is the timestamp for the occurrence of the real Near state, while $t_{sys}$ is the timestamp of the first system detection.
    
    \item False alarm (FA) rate. When the proposed system mistakenly detects a subject enters the target area, a false alarm happens, as indicated by Fig. \ref{fig:false_alarm_metric}. The false alarm rate, also instance-based, can be defined as
    \begin{equation}
        FA = \frac{N^{FA}_{sys}}{N^{empty}_{GT}},
    \end{equation}
    where $N^{FA}_{sys}$ is the number of false detection events, and the denominator $N^{empty}_{GT}$ corresponds to the total number of instances where the target area is empty.
\end{itemize}

\subsection{Experimental Results}
\subsubsection{Performance in different environments}
The performance in the four environments is shown in Table \ref{tab:performance}. The parameters are determined empirically and universal for all environments. 
The overall IA and FA rate are 92.5\% and 1.12\%, respectively. As we can see, the results demonstrate a high IA of $>$97\% and a low FA rate of $<$3\% in all environments, highlighting the effectiveness of the proposed approach in accurately detecting proximity in various scenarios. Also, once a proximity instance is detected, the system also exhibits a near-perfect DA of 98.86\%, aligning consistently with the ground truth. The overall detection delay of 0.825s also demonstrates a responsive system detection. 

\begin{table}[t]
    \centering
    \begin{adjustbox}{width=\linewidth,center}
    \begin{tabular}{c|c|c|c|c|c|c}
    \hline
          Environment & Samples$_{IA}$ & $IA$ & $DA$ & $\tau$ & Samples$_{FA}$ & $FA$ \\
          \hline
          Office 1 & 83 & 91.6\% & 98.4\% & 0.65s & 72 & 1.39\%\\
          \hline
          Office 2 & 46 & 93.5\% & 97.2\% & 0.71s & 62 & 0.0\%\\
          \hline
          Home 1 & 56 & 92.9\% & 99.5\% & 1.12s & 60 & 0.0\%\\
          \hline
          Home 2 & 68 & 92.6\% & 100\% & 0.87s & 75 & 2.67\%\\
          \hline
          Overall & 253 & 92.5\% & 98.86\% & 0.825s & 269 & 1.12\%\\
     \hline
    \end{tabular}
    \end{adjustbox}
    \caption{Performance in different environments.}
    \label{tab:performance}
\end{table}

\subsubsection{The impact of walking path length}

\begin{figure}[t]
  \centering
  \includegraphics[width=0.9\linewidth]{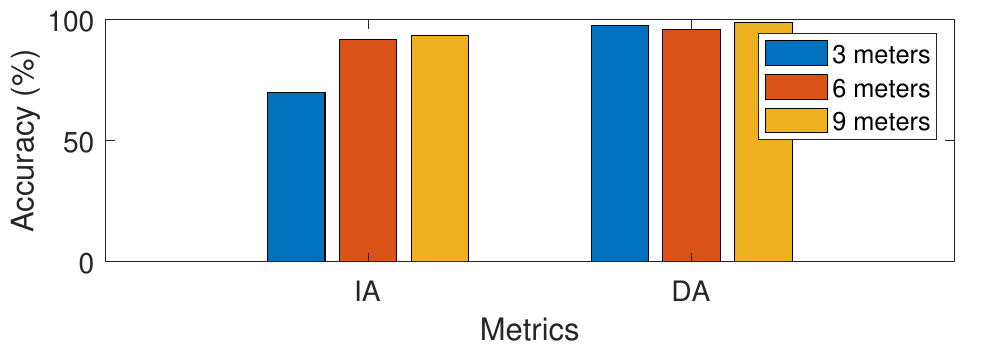}
    \caption{Impact of length of walking path.}
    \label{fig:walking_path}
\end{figure}

Although the previous section has demonstrated reliable detection performance in different environments, the tested walking paths are usually longer than 5 meters to ensure a stable walking period. To evaluate how the length of the walking path affects the performance, we collect more data with different walking path lengths in Environment 1, including 3 meters, 6 meters and 9 meters by adjusting the starting points. The accuracy shown in Fig. \ref{fig:walking_path} indicates that the shorter the path length, the lower the IA, as the gait pattern may not be stable enough during the transition period. Since the DA is calculated based on the successful recognition of the approaching and leaving states, as seen from the Fig. \ref{fig:walking_path}, when the approaching and leaving states are identified, the DA value approaches near 100\%.

\section{Discussions and Future Work}
\label{sec:future_work}
In this section, we discuss the limitations of the proposed  algorithm and explore potential future directions. First, the proposed proximity detection algorithm performs well assuming normal walking from a single human subject. In practice, various users have different walking styles, requiring careful parameter calibration. 
To address this limitation, future research could focus on developing adaptive algorithms that can automatically adjust the parameters based on the specific user's walking style. This would enhance the algorithm's robustness and ensure accurate proximity detection across different individuals.

Second, due to channel reciprocity, the proposed system exhibits similar responses as the user approaches the Tx and the target IoT devices, i.e., the Rx in our experiments. To eliminate this  ambiguity, we propose deploying additional IoT devices and jointly observing their reactions. If all/most the Rx devices show correlated features/transitions, it indicates a movement around the Tx; otherwise, it suggests that the user is near a specific Rx.

\section{Conclusion}
\label{sec:conclusion}

In this paper, we propose a novel approach of using the gait analysis to improve the proximity detection. By integrating the gait score with the proximity feature and employing a state machine, we achieve precise and continuously stable detection.
Extensive experiments validates the effectiveness of our system, with high detection rates and low false alarms. Our approach will benefit applications in security systems, healthcare monitoring, and smart environments, with opportunities for future refinement and real-world implementation.

\bibliographystyle{ieeetr}
\bibliography{refs}


\end{document}